\documentclass[12pt]{article}
\usepackage{amssymb,amsmath,amsthm,graphicx,ulem}

\pdfoutput=1

\usepackage{graphicx,subfigure}
\usepackage{epsfig}
\usepackage{amsmath}
\usepackage{amsfonts}
\usepackage{amssymb}
\usepackage[usenames]{color}
\usepackage[letterpaper,left=2.cm,right=2.cm,top=2.5cm,bottom=2.5cm]{geometry}
\usepackage[T1]{fontenc}

\newcommand{\beq}{\begin{equation}}
\newcommand{\eeq}{\end{equation}}
\newcommand{\be}{\begin{equation}}
\newcommand{\ee}{\end{equation}}
\newcommand{\beqa}{\begin{eqnarray}}
\newcommand{\eeqa}{\end{eqnarray}}
\newcommand{\beqar}{\begin{eqnarray*}}
\newcommand{\eeqar}{\end{eqnarray*}}
\newcommand{\bea}{\begin{eqnarray}}
\newcommand{\eea}{\end{eqnarray}}

\numberwithin{equation}{section}

\newcommand{\nn}\nonumber
\newcommand{\eqn}[1]{(\ref{#1})}

\numberwithin{equation}{section}

\begin{document}

\allowdisplaybreaks

\normalem

\title{Entanglement Entropy Near Cosmological Singularities}

\author{Netta Engelhardt and Gary T. Horowitz
\\ 
\\ \\
  Department of Physics, UCSB, Santa Barbara, CA 93106, USA \\ 
 \\ 
 \small{engeln@physics.ucsb.edu, gary@physics.ucsb.edu}}

 \date{}

 \maketitle

\begin{abstract}
\noindent We investigate the behavior of the entanglement entropy of a confining gauge theory near cosmological singularities using gauge/gravity duality.  As expected, the coefficients of the UV divergent terms are given by simple geometric properties of the entangling surface in the time-dependent background. The finite (universal) part of the entanglement entropy either  grows without bound or remains bounded depending on the nature of the singularity and entangling region. We also discuss a confinement/deconfinement phase transition as signaled by the entanglement entropy.

\end{abstract}

\newpage


\tableofcontents
\baselineskip16pt

\section{Introduction}
\noindent There has been considerable interest recently in the entanglement entropy of quantum field theories. Entanglement entropy is a measure of long range correlations in the system which has proven useful in a variety of applications including  identifying exotic ground states  \cite{KitaevPreskill, LevinWen} and signaling phase transitions. In particular, it has been shown \cite{Nishioka:2006gr,Klebanov:2007ws,RefinedHEE} that a confinement/deconfinement phase transition, which is usually studied at finite temperature,  can be detected in the zero temperature ground state by studying the entanglement entropy for regions of various sizes.

We explore another potential application of the entanglement entropy: probing cosmological singularities. Understanding past spacelike (``big bang'') singularities is a long-standing goal of quantum gravity. While much progress has been made in understanding static timelike singularities in string theory, spacelike singularities  remain mysterious. There have been various attempts to resolve spacelike singularities using gauge/gravity duality (see, e.g.,  \cite{Craps:2005wd,HertogHorowitz,Turok:2007ry,BurgessMcAllister}) but we will be less ambitious here. Rather than attempt to resolve cosmological singularities, we will endeavor to understand one of their effects on a quantum field theory. We compute the entanglement entropy for a confining gauge theory in a class of fixed (classical) cosmological backgrounds. As we will discuss, the choice of a confining gauge theory has technical advantages for the calculation, but it is also of particular physical interest since it
models an early-universe quark-gluon plasma. The quark-gluon plasma is expected to feature a mass gap and a confinement/deconfinement phase transition at around 150 MeV \cite{Satz:2008kb}.

Given a state of a quantum field theory and a region of space, $\mathcal{R}$, the entanglement entropy  measures  the quantum entanglement  of the part of the state in $\mathcal{R}$ with the rest of the state in $\overline{\mathcal{R}}$, the complement of $\mathcal{R}$. It is defined in terms of the von Neumann entropy of the reduced density matrix on $\mathcal{R}$:
\be S_{EE} = - \rm{Tr}\left (\rho_{\mathcal{R}} \ln \rho_{\mathcal{R}}\right) \label{vonNeumann}\ee 
\noindent where $\rho_{\mathcal{R}} = \rm{tr}_{\overline{\mathcal{R}}}\rho$.  While $S_{EE}$ is a divergent quantity due to the short-wavelength modes across $\partial \mathcal{R}$ and must be regulated, it is generally possible to extract its universal (UV cutoff-independent) component. 

Calculating $S_{EE}$ in strongly coupled theories using the field theory formalism, however, is notoriously difficult. We will therefore calculate $S_{EE}$ using the more tractable holographic approach instead. The original Ryu-Takayanagi  formulation \cite{Ryu:2006bv} of holographic entanglement entropy states that the entanglement entropy of a region $\mathcal{R}$ in a  field theory dual to some static asymptotically locally AdS$_{d+1}$ bulk spacetime is proportional to the area of a minimal  surface homologous to the entangling region on the boundary:
\be S_{EE} = \frac{\mathrm{Area} (\gamma_{\mathcal{R}})}{4 G_{N}^{(d+1)}} \label{RT} \ee
\noindent where $\gamma_{\mathcal{R}}$ is the minimal area surface in the bulk and $G_{N}^{(d+1)}$ is the $(d+1)$-dimensional Newton's constant\footnote{Note that this proposal admits quantum and stringy corrections when the bulk spacetime is not taken to obey classical general relativity.}. In this proposal, the classical bulk spacetime is static, so there is a canonical time slicing; the region $\mathcal{R}$ and the corresponding minimal surface $\gamma_{\mathcal{R}}$ both lie on the same constant time slice. The more general, covariant Hubeny-Rangamani-Takayanagi  prescription calls for the minimal surface $\gamma_{\mathcal{R}}$ to be replaced by an extremal surface; if several exist, the one with minimal area is chosen \cite{HRT}. Since we work in cosmological backgrounds, we will employ the covariant proposal here.

There have been earlier studies of entanglement entropy in time-dependent backgrounds, e.g., \cite{AbajoArrastia:2010yt,Aparicio:2011zy,Albash:2010mv,Balasubramanian:2010ce,AsplundAvery,Basu:2011ft,Basu:2012gg,Maldacena:2012xp,Buchel:2013lla,Nozaki:2013wia}, but to our knowledge, none have explored the behavior near singularities.\footnote{The interesting recent paper \cite{Hartman:2013qma} considered the contribution to entanglement from the region inside black holes, but the extremal surfaces were bounded away from the singularity in the bulk.}

\indent The area of any bulk surface which extends all the way to the boundary at infinity diverges, and this divergence obeys  \cite{RT09}:
\be\mathrm{ Area}(\gamma_{R}) = \frac{c_{d-2}}{\epsilon^{d-2}} + \frac{c_{d-4}}{\epsilon^{d-4} } + \cdots +\left\{ 
  \begin{array}{l l}
c_{0} \log \left (\frac{L}{\epsilon} \right) + \cdots & \text{$d$ is even} \\
c_{0} +\cdots & \text{$d$ is odd}
  \end{array} \right.  \label{arealaw} \ee
\noindent where $L$ is the length characterizing $\mathcal{R}$, $\epsilon$ is a UV cutoff, and the dots in the brackets represent terms that vanish as we take $\epsilon$ to zero. If a boundary-covariant cutoff is used, the coefficients $c_i$ of the divergent terms are integrals of geometric quantities. In particular, the leading coefficient is proportional to the area of the boundary of the entangling region, $\partial \mathcal{R}$. We will work with $d=5$ and determine the $c_1$ coefficient as well as the UV cutoff-independent $c_0$ as functions of time in a class of cosmological backgrounds.

We  treat the bulk geometry classically, but expect stringy and quantum corrections to be important very close to the singularity. We will stay away from this poorly understood region and only consider the entanglement entropy for times where the extremal surface stays in a region of the bulk where classical general relativity is valid.

\section{Cosmological backgrounds and extremal surfaces}
\noindent We wish to calculate the entanglement entropy for a field theory in a Kasner$_{3+1}\times S^{1}$ background:
\begin{equation} ds_{\text{bdy}}^{2}= -dt^{2} + t^{2p_{1}}dx^{2} + t^{2p_{2} }dy_{1}^{2} + t^{2p_{3}}dy_{2}^{2} + d\theta^{2}\label{BoundaryMetric} \end{equation}
\noindent where three of the spatial directions  are periodically identified:
\begin{align*}  & \theta  \sim \theta + L_{0} \\
& y_{1}  \sim y_{1} +L_{1} \\
& y_{2}  \sim y_{2} +L_{2} \end{align*}
\noindent and the exponents $p_{i}$ obey the Kasner conditions:
\begin{equation}\label{sumrule}
 \sum\limits_{i} p_{i} = 1 = \sum\limits_{i} p_{i}^{2}
  \end{equation}
These last conditions ensure that the metric is Ricci flat. This spacetime describes a homogeneous but anisotropic cosmology and generically has a curvature singularity at $t=0$. The constant $t$ surfaces are flat, but space expands at different rates in different directions.  Since there are two conditions on the three exponents $p_i$, there is a one parameter family of solutions  labeled, e.g., by $p_1$ which can range from $-1/3$ to $1$. Generically at least one $p_i$ is negative and the rest are positive, so one direction contracts while others  expand. The only exception is when two of the exponents vanish and one recovers  the Milne solution, which is locally flat. This is the only case in which $t=0$ is not a curvature singularity.

We choose our entangling region, denoted $\mathcal{R}(t_{b},x_{0})$,  to be a strip which is extended  in all spatial directions except one, on a fixed time slice $t=t_{b}$:
\be \mathcal{R}(t_{b},x_{0}) = 
 \left\{ 
  \begin{array}{l l}
& x  \in [-x_{0}, x_{0}]\\
& \theta  \in [0, L_0)\\
& y_{1}  \in [0, L_{1})\\
& y_{2}  \in [0, L_{2})\\
& t  = t_{b}
  \end{array} \right. \label{EntanglingRegion}\ee
\noindent It is the goal of this paper to calculate the entanglement entropy of $\mathcal{R}(t_{b},x_{0})$ holographically as a function of $t_{b}$. To do so, we must first find a bulk dual to the Kasner$_{3+1}\times S^{1}$ background.

\indent We start with the AdS$_6$ soliton metric \cite{Witten:1998zw,HorowitzMyers}, obtained by analytically continuing the timelike direction and one spatial direction from the planar AdS black hole:
\begin{equation}  ds_{\text{soliton}}^{2} = \frac{1}{z^{2}} \left [ \left(1-z^{5}\right) d\theta^{2} + \eta_{\mu\nu} dx^{\mu}dx^{\nu} + \frac{dz^{2}}{1-z^{5}} \right] \label{SolitonMetric}\end{equation}
\noindent where  the AdS radius and black hole horizon radius have both been set to one. The $\theta$ circle smoothly caps off at $z=1$, provided we set $L_0 =  \frac{4 \pi}{5}$, i.e., we periodically identify $\theta \sim \frac{4 \pi}{5}+\theta$.  This solution describes a confining vacuum in the dual field theory which lives on $\mathbb{R}^{3,1}\times S^{1}$. One can change the boundary metric by using the fact that  replacing $\eta_{\mu\nu}$ in (\ref{SolitonMetric}) with any $(3,1)$-dimensional, Ricci-flat solution of Einstein's equations  still results in a bulk metric which is an exact solution of Einstein's equations with a negative cosmological constant.  Replacing $\eta_{\mu\nu}$ with Kasner yields the $(5+1)$-dimensional Kasner-AdS soliton:
\begin{equation} ds_{\text{KAS}}^{2}= \frac{1}{z^{2}}\left ( \frac{dz^{2}}{1- z^{5}}-dt^{2} + t^{2p_{1}}dx^{2} + t^{2p_{2}} dy_{1}^{2} + t^{2p_{3}}dy_{2}^{2} +(1-z^{5}) d\theta^{2} \right)\label{Kasneton} \end{equation}
 This solution has a spacelike singularity at $t=0$ which extends across the entire spacetime.

We could have started with AdS in Poincar\'e coordinates, and replaced the Minkowski metric on each constant radial surface with Kasner. The result is again a solution of Einstein's equation with negative cosmological constant. This would be dual to a nonconfining gauge theory on the Kasner spacetime.  However, in addition to the singularity at $t=0$, this solution has singularities on the Poincar\'e horizon which appear to be unphysical. These unphysical singularities are removed in \eqn{Kasneton} since the radial direction is capped off and there is no Poincar\'e horizon. Another advantage of starting with the AdS soliton is
that since it is a confining geometry, it can be used to probe the confinement/deconfinement phase transition.

 The proposal in \cite{HRT} dictates that entanglement entropy be calculated from the area of an extremal surface homologous to the entangling region. We use the symmetries of the surface and the bulk metric to parametrize the surface in terms of the distinguished boundary coordinate $x$:
	\begin{equation} X^{\mu}=\left (t,x,\theta, y_{1},y_{2},z\right ) = \left ( T(x), x, \theta, y_{1},y_{2},Z(x)\right) \label{ExtremalSurface}\end{equation}
\noindent The induced metric on the surface $\gamma_{ab} = \partial_{a}X^{\mu}\partial_{b}X^{\nu}g_{\mu \nu}$, where $g_{\mu \nu}$ is the bulk metric, gives rise to the area functional for \eqn{ExtremalSurface}:
\begin{equation} A= \frac{4 \pi L_1 L_2}{5} \int\limits^{x_{0}}_{0} \ dx \frac{T(x)^{1-p_{1}}}{Z(x)^{4}} \left [\left(1-Z(x)^{5}\right)\left( T(x) ^{2p_{1}} -T'(x)^{2}\right) + Z'(x)^{2}\right]^{\frac{1}{2}}\label{UnregulatedArea} \end{equation}
\noindent The set $\{T\left(x\right),Z\left (x\right)\}$ which solves the equations of motion generated by \eqn{UnregulatedArea} and obeys the following boundary conditions
\be
	\begin{aligned}
	& Z\left(x=\pm x_{0}\right)=0\\
	&  T\left(x=\pm x_{0}\right)=t_{b}
	\end{aligned}
\label{BdyConditions}
\ee
describes the extremal surface that enters into \eqn{RT}. \\
\indent We pause here to note that the integral in \eqn{UnregulatedArea} diverges and must be regulated. It is conventional to impose a UV cutoff some small distance away from the boundary $z=\epsilon$, which is the same UV cutoff that appears in the expansion \eqn{arealaw}. The integral in \eqn{UnregulatedArea}, however, is over the variable $x$, so we define the UV cutoff  to be the value $z=\epsilon$ such that $Z(x_{0}-\delta)=\epsilon$. We then integrate up to $x=x_{0}-\delta$ instead of up to $x_{0}$. \\

\subsection{Finding the Extremal Surfaces}

\noindent The equations of motion generated by \eqn{UnregulatedArea} are a set of two coupled second order ordinary differential equations for which an analytical solution is not known. The general solution depends on four parameters, but the local boundary conditions \eqn{BdyConditions} reduce this to two. We will generate a 2-parameter family of approximate solutions to the equations of motion by truncating a set of series expansions, and then use numerics to determine those two parameters for different values of the power $p_{1}$ and the time slice of the entangling region $t_{b}$. We now describe this procedure in more detail. Since it is valid in arbitrary dimension, in this subsection we work in general $d$.

\indent We first solve the equations of motion for the surface via power series expansions. Integer power series, however, cannot represent nonanalytic functions. Since extremal surfaces in pure AdS are nonanalytic at the boundary and the Kasner-AdS soliton is asymptotically locally AdS, we expect extremal surfaces in \eqn{Kasneton} to have the same nonanalytic asymptotic behavior. We therefore investigate the asymptotic behavior of $T(x)$ and $Z(x)$ to find the corresponding fractional power series.\\
 \indent To probe the asymptotic behavior of $T(x)$, consider the AdS$_{d+1}$ soliton metric \eqn{SolitonMetric}  in Milne coordinates:
\be ds^{2} = \frac{1}{z^{2}} \left (\left( 1- z^{d} \right)d\theta^{2} + \frac{dz^{2}}{1-z^{d}}  -dt^{2} + t^{2} dx^{2}+ dy_{1}^{2} + dy_{2}^{2} \right)\label{MilneSoliton}\ee
\noindent Implementing a simple coordinate redefinition, we obtain the manifestly static soliton:
\begin{equation} 
		\begin{aligned}
		\chi & = t \sinh x\\
		\tau &= t \cosh x 
		\label{MilneTransform}
		\end{aligned}
\end{equation} 
\be ds^{2} = \frac{1}{z^{2}} \left (\left( 1- z^{d} \right)d\theta^{2} + \frac{dz^{2}}{1-z^{d}}  -d\tau^{2} + d\chi^{2}+ dy_{1}^{2} + dy_{2}^{2} \right)\label{SolitonTransform}\ee
\noindent The boundary condition $T(x=\pm x_{0})=t_{b}$ in the Milne soliton corresponds to a boundary condition $\tau(x =\pm x_{0}) = t_{b} \cosh x_{0}$ in the soliton. The soliton, however, is static, so the entire extremal surface rests on the time slice $\tau = t_{b} \cosh x_{0}$. Using \eqn{MilneTransform}, we obtain an equation for $T(x)$:
\be T_{\text{Milne}}(x) = t_{b} \cosh x_{0} \text{sech} \ x \ee
\noindent $T(x)$ is therefore analytic in the Milne soliton. We expect that, for the  Kasner-AdS soliton, $T(x)$ is analytic as well, and can be expanded:
\be T(x)= \sum\limits_{i=0}^{\infty} T_{i}\left (x_{0}^{2}-x^{2}\right )^{i}\label{TAsymptoticForm}\ee
\noindent where the $x^{2}$ term results from $T(x)=T(-x)$ by reflection symmetry of the entangling region. \\
\indent The exact solution for $x(Z)$ is known in pure AdS \cite{Hubeny}:
\begin{equation}\pm x(Z) = \frac{Z^{d}}{d z_{*}^{d-1}}\ _{2}F_{1} \left [ \frac{1}{2}, \frac{d}{2(d-1)}, \frac{3d-2}{2(d-1)}, \frac{Z^{2(d-1)}}{z_{*}^{2(d-1)}}\right]- z_{*} \frac{\sqrt{\pi} \Gamma \left [ \frac{3d-2}{2(d-1)}\right]}{d \Gamma \left [ \frac{2d-1}{2(d-1)}\right]}\end{equation}
\noindent where 
\be
z_{*} = 2x_{0} \frac{(d-1) \Gamma \left [ \frac{2d-1}{2(d-1)}\right]}{\sqrt{\pi} \Gamma \left [ \frac{d}{2(d-1)}\right]}
\ee
 is the turning point of the surface and $_{2}F_{1}$ is a hypergeometric function. We expand the solution in a Taylor series about the point $x_{0}$ and then take the inverse series to obtain an expansion for $Z(x)$:
\begin{equation} Z(x) = (x_{0}^{2}-x^{2})^{1/d} \sum\limits_{i=0}^{\infty} Z_{i} (x_{0}^{2}-x^{2})^{\frac{i}{d}}\label{ZAsymptoticForm}\end{equation}
\noindent where the $Z_{i}$ are constants. We expect a similar asymptotic behavior in the Kasner-AdS soliton. Indeed, 
substituting the expansions \eqn{TAsymptoticForm} and \eqn{ZAsymptoticForm} into the equations for the extremal surface results in a two parameter family of solutions. The remaining two parameters are fixed by requiring regularity at the origin. This is difficult to impose starting with local power series expansions about $\pm x_0$, so we turn to numerical methods.

One can make the expansion for $Z(x)$ analytic by introducing a new coordinate $\tilde{x}$ defined by
$\left (x_{0}^{2}-x^{2}\right)^{\frac{1}{d}}= \left ( 1- \tilde{x}^{2}\right)$.
The equations of motion remain analytic in $\tilde{x}$ and we have
\be 
Z(\tilde x) = \left ( 1- \tilde{x}^{2}\right) \sum\limits^{\infty}_{i=0} \tilde{Z}_{i} \left ( 1- \tilde{x}^{2}\right)^{i}  = \left ( 1-\tilde{x}^{2}\right) P\left (\tilde{x}\right)
\ee
The factor of $\left ( 1-\tilde{x}^{2}\right)$ on the right is convenient since then $P(\pm 1)$ is nonzero. 
Since $P(\tilde{x})$ is analytic everywhere, we can solve for it and $T(\tilde x)$ numerically using pseudospectral methods. For a single connected extremal surface, this is most conveniently  done by taking advantage of the reflection symmetry about $x=0$ and just  solving the equations for $x$ in $[0,x_0]$ with
boundary conditions $T'(0) = 0, Z'(0) = 0, T(x_0) = t_b$ and $Z(x_0) = 0$.
We have checked our numerics by comparing the numerical solutions with known exact solutions in special cases. Our numerical solutions for $Z(x)$ in pure AdS and $T(x)$ in the Milne soliton agree with the known solutions  to order $10^{-14}$. We solve the equations of motion for a range of the parameters $p_{1}$ and $t_{b}$. By comparing the power series solutions with the numerical results, we fix the two free parameters in those solutions. 

 The reason that we choose to work with analytical power series rather than the numerical solutions is that numerical integration of the surfaces is difficult to perform accurately due to the steepness of the surfaces at boundary. In particular, extracting the dependence of the area on the cutoff is delicate to do numerically, and much more accurate results can be obtained from the power series as follows.
We compute the area by substituting the (truncated) series expansions for $T(x)$ and $Z(x)$ into \eqn{UnregulatedArea}, expand the functional in polynomials, and integrate analytically from $x=0$ to $x=x_{UV}$, where $x_{UV}$ is a symbolic parameter without an assigned numerical value. The analytical form of the integral $A(x)$ is a power series in $x_{UV}$:
\be A(x_{UV}) = \sum\limits_{i}A_{i} \left(x_{0} - x_{UV}\right)^{\frac{i}{d}}\label{A(x)} \ee
\noindent where the $A_{i}$ are numerical coefficients. In order to determine the area's  UV cutoff-dependent and independent behavior, the area $A(x_{UV})$ must be expressed as a function of $Z$. To accomplish this, we obtain a series expansion for $x(Z)$ by inverting the series $Z(x)$ via a variable change $\left (x_{0}^{2} - x^{2}\right)^{1/d} = x_{0}^{2} - y^{2}$, so that we have, as before, an analytic series expansion for $Z$: 
\be Z(y) = \sum\limits_{i}\mathcal{Z}_{i} \left( x_{0}^{2} -y^{2}\right)^{i} \ee
\noindent Since this series is analytic, we can easily obtain the inverse series $y(Z)$ and the area $A\left (x\left ( y \left (Z\right) \right)\right)=A\left (Z\right )$ as a function of the bulk coordinate $Z$. A Taylor expansion of $A(Z)$ around $Z=0$ yields the expected behavior \eqn{arealaw}. This allows us to immediately read off the leading divergent, subdivergent, and constant pieces for a choice of pair $(p_{1},t_{b})$. This process is repeated for a representative range of $p_{1}$ and $t_{b}$. We use a least-squares best-fit algorithm to obtain the  coefficients of the divergent, subdivergent, and UV cutoff-independent pieces as functions of $t_{b}$ and $p_{1}$.

\section{Results \label{sec:Results}}
\subsection{Confinement/deconfinement transition}

 We first describe the behavior of the extremal surfaces in the bulk.\footnote{The specific calculations described in this section were done with $x_0 = 0.12$. This is a convenient value since it is large enough that the extremal surfaces do not stay just in the asymptotic AdS region, and small enough that the extremal surfaces stay connected for a range of $t_{b}$.
  It is worth noting that while $x_{0}=0.12$ is a convenient choice for the numerical analysis, the qualitative behavior is independent of this choice.} The possibility that extremal surfaces will bend toward the singularity and reach it at some nonzero value of the boundary time $t_b$ may  appear to be a concern, as this would make it impossible to evaluate the entanglement entropy at any earlier time. We find that this does not occur. The bulk time $t$ increases along the extremal surface, reaching a maximum $t_*$ at the midpoint $x=0$. For $t_b$ of order one, the increase is small (of order a few percent), but closer to the singularity it becomes larger. For example, for $p_1= -0.25$ and $t_b = 0.4$, $t_*$ is $20\%$ larger than $t_b$.

Next we consider the radial extent of the extremal surface. For a static strip in pure AdS$_6$, the extremal surface reaches a maximum value of $z$ given by \cite{Hubeny}
\be
 z_* = \frac{4\Gamma(9/8)}{\sqrt \pi \Gamma(5/8)}\mathcal{L} \approx 1.48 \mathcal{L}\label{zstar}
 \ee
where $\mathcal{L}$ is the width of the strip. We find that this formula works remarkably well for the Kasner-AdS soliton with $\mathcal{L}$ interpreted as the proper width of the entangling region, until $z_*$ gets close to one. The fact that the cap in the radial direction due to the soliton does not  affect the surface much when $z_* < 1$ can be understood from the fact that we work in six dimensions, so the effect of the cap comes in at $O(z^5)$ in \eqn{Kasneton}. It is more surprising that the time dependence does not seem to affect the radial extent of the extremal surface (at least in the coordinates used in \eqn{Kasneton}). Eq. \eqn{zstar} holds even when the extremal surface does not stay near a constant $t$ surface in the bulk.

\begin{figure}
\centerline{
\includegraphics[width=0.3\textwidth]{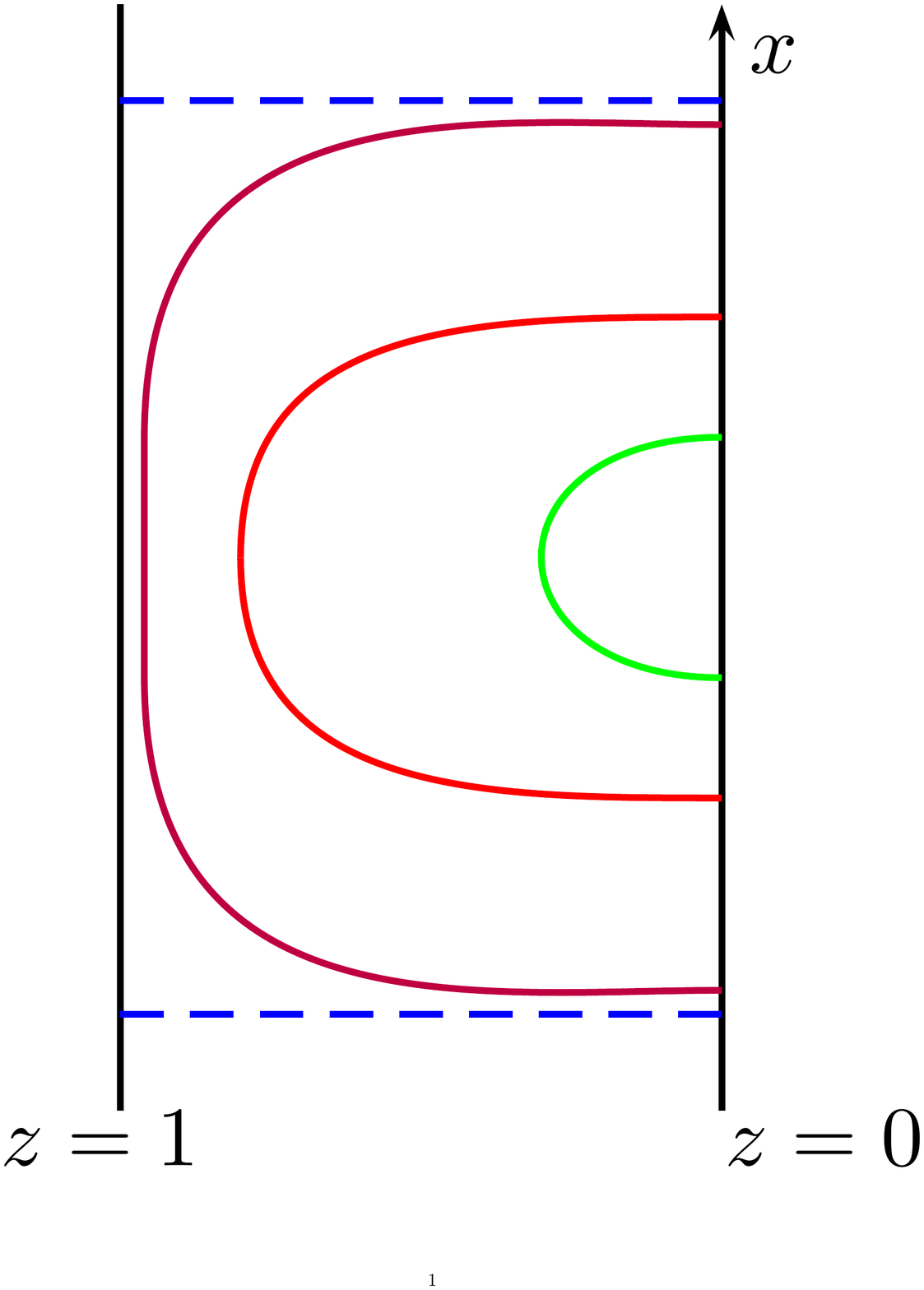}
}
\caption{\small For small proper width $\mathcal{L}$ in the Kasner-AdS soliton background, the extremal surfaces remain close to the boundary and resemble surfaces in pure AdS. The area of the surface depends on $\mathcal{L}$, so the entanglement entropy is a function of $\mathcal{L}$ and the modes which contribute are deconfined.  As the length of the entangling region becomes larger, extremal surfaces extend further into the bulk. The soliton cap begins to affect their geometry, and they start leveling out. For some critical length, the surface splits into two separate surfaces (displayed above in dashed blue). At this point, the area of the extremal surface is independent of $\mathcal{L}$, effectively signaling that $1/\mathcal{L}$ is below the mass gap of the confined gauge theory. }
\label{fig:SurfaceTransition}
\end{figure}

When $z_*$ does approach one, there is a qualitative change in the extremal surface. Rather than one surface connecting the two ends of the strip, the extremal surface splits into two pieces, with each one wrapping the cap at $z=1$ (see Fig.~\ref{fig:SurfaceTransition}).  The topology of the extremal surface in the $x, z,\theta$ directions changes from $\mathbb{R}\times S^1$ to two copies of $\mathbb{R}^2$. Once the extremal surface splits, the entanglement entropy becomes independent of the width $\mathcal{L}$.

This transition is the holographic description of a confinement/deconfinement phase transition \cite{Nishioka:2006gr,Klebanov:2007ws}. Entanglement entropy of a confining gauge theory measures the effective degrees of freedom between the entangling region and its complement at the energy scale $E \sim \frac{1}{\mathcal{L}}$.  The corresponding modes of the gauge theory are in the deconfined phase when $E$ is large and the entanglement entropy is a function of $\mathcal{L}$. As $\mathcal{L}$ grows, one probes progressively lower energy modes. Eventually one hits the mass gap for the confining theory and there are no modes of lower energy. When this happens, the entanglement entropy becomes independent of $\mathcal{L}$. 

\begin{figure}
\centering

\subfigure[ ]{
\centering
\includegraphics[width=0.43\textwidth]{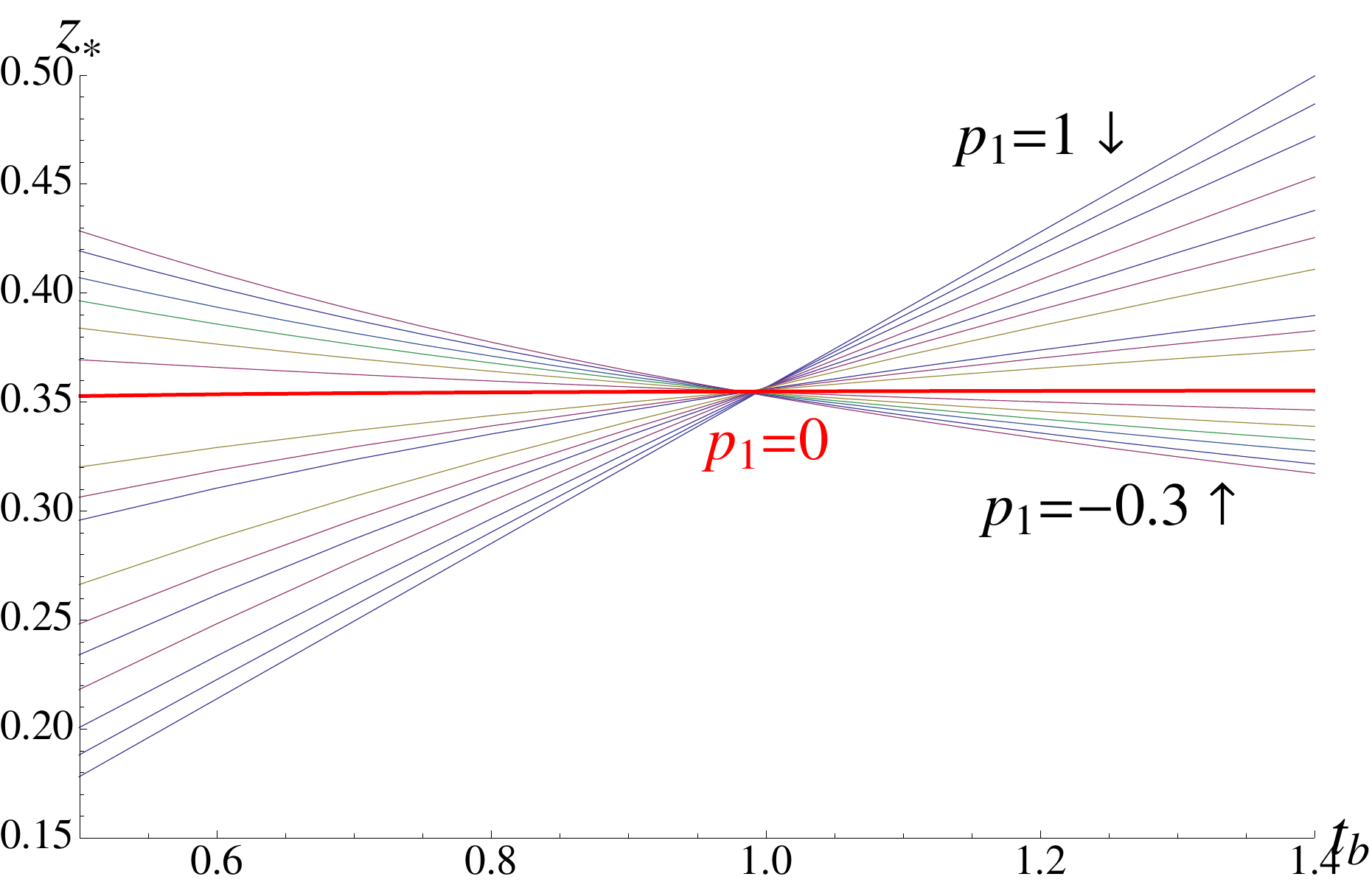}
}
\hfill
\subfigure[]{
\centering \includegraphics[width=0.43 \textwidth]{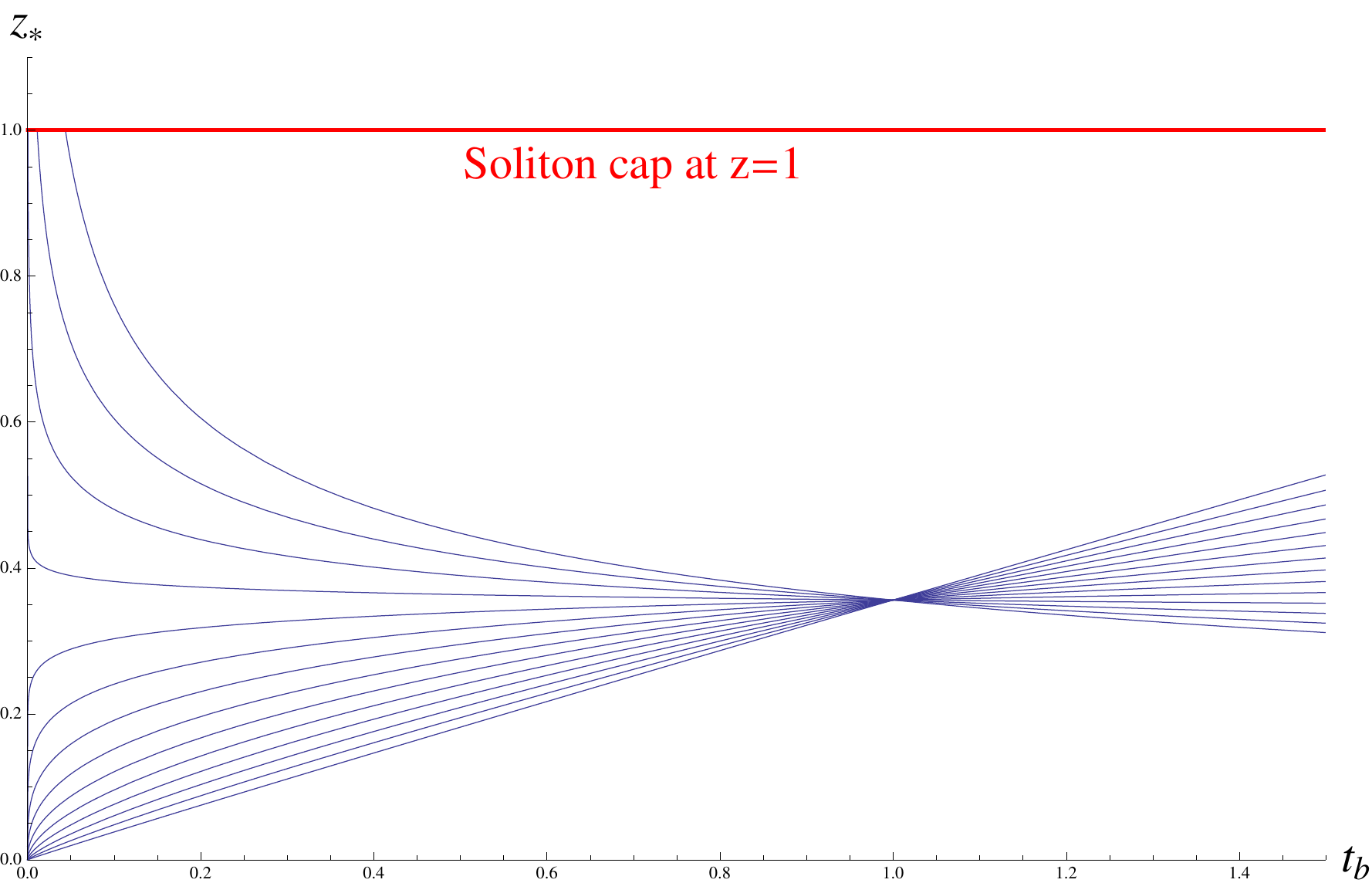}
}
\hfill
\caption{ (a): \small The maximum value of $z$ as a function of $t_{b}$, the boundary time. Each line corresponds to a different value of $p_{1} \in [-0.3, 1]$: surfaces with positive $p_{1}$ shrink towards the boundary as $t_{b}\rightarrow 0$ while surfaces with $p_{1}<0$ approach the cap. Surfaces with $p_{1}=0$ maintain constant $z_{*}$. (b): A plot of $z_{*}(t_{b})= 1.48 \mathcal{L}$ for small $t_{b}$ and $p_{1} \in [- \frac{1}{3}, 1]$. This describes the radial extent of minimal surfaces until $z_* \approx 1$ when the surface splits into two, signaling a confinement/deconfinement transition. }
\label{fig:ProjectedTransition}
\end{figure}

The sign of $p_1$ clearly plays an important role in determining when this transition occurs. When $p_1<0$, the proper length $\mathcal{L}$ grows as one approaches the singularity and the confinement/deconfinement transition takes place at early time. When $p_1 >0$,  $\mathcal{L}$ grows as one moves away from the singularity, so the confinement/deconfinement transition occurs at late time (see Fig.~\ref{fig:ProjectedTransition}). 

\subsection{Entanglement entropy}
\indent We now proceed to present the behavior of the area of the extremal surface as a function of $t_{b}$ and $p_{1}$. As noted above, this is divergent, so we introduce a small cutoff at $Z=\epsilon$ and expand  $A(Z=\epsilon)$. It is convenient to factor out the trivial integral over $y_1,y_2,\theta$ and write: 
\begin{equation} \frac{A(\epsilon)}{L_{1} L_{2} \frac{4\pi}{5}}= \frac{c_{3}}{\epsilon^{3}} + \frac{c_{1}}{\epsilon} + c_{0} + \cdots \end{equation}
\noindent As expected, the subdivergent piece $1/\epsilon^{2}$ vanishes, and the leading coefficient is proportional to the area of the entangling region:
\begin{equation} c_{3} =\frac{1}{3} \frac{A_{\text{bdy}}}{L_{1} L_{2}\frac{4\pi}{5} } =\frac{2}{3} t_{b}^{p_{2}+p_{3}} = \frac{2}{3}t_{b}^{1-p_{1}}\end{equation}
\noindent where the factor of 2 is due to the fact that the $y_{i}$ and $\theta$ are maximally extended, so $\partial \mathcal{R}$ consists of two identical disconnected pieces. The last equality comes from the Kasner conditions \eqn{sumrule}.
The factor of $1/3$ in the first equality can be understood geometrically: the bulk is asymptotically locally AdS$_6$, so near $Z=0$, the area functional is approximately: 
\begin{equation} A_{\text{AdS}}= \int \frac{A_{\text{bdy}}}{Z^{4}}dZ\end{equation}
\noindent The $Z^{4}$ factor results in a $1/3$ contribution to the integral. This result moreover agrees with \cite{RefinedHEE}, where the computation is done for the pure soliton for an entangling region identical to $\mathcal{R}\left (t_{b},x_{0}\right)$.

\indent We next turn to the subleading divergence. We find $c_1(t_{b})$ is a simple power law:
\begin{equation}
	\begin{aligned} c_{1}(t_{b})  & =\gamma \left (p_{1}\right) t_{b}^{ \beta \left (p_{1}  \right)}\\
				\gamma\left(p_{1}\right) & = a +b p_{1} +cp_{1}^{2} \\
				\beta \left(p_{1}\right) & = -1-p_{1} \ 
	\end{aligned}
\label{c1}
\end{equation} 
\noindent where $a, \ b,$ and $c$ are numerical constants. 

Considerations of covariance and dimensional analysis lead us to conclude that the $c_{1}$ coefficient is an integral of a dimension two geometric quantity. This can be  the scalar curvature of the background spacetime, the induced metric on the boundary of the entangling region, or a scalar constructed from the  square of the extrinsic curvature, ${K_{\mu\nu}}^{\rho} $. Both scalar curvatures vanish in our case, so the two possible contributions are:
\begin{equation} A_{\text{bdy}} \left ( \gamma_{1} K_{\mu}^{\mu} \ _{\rho} K^{\nu} _{\nu} \ ^{\rho}  + \gamma_{2} K_{\mu \nu}\ ^{\rho}K^{\mu \nu} \ _{\rho}   \right)= - \frac{A_{\text{bdy}}}{t_{b}^{2}}\left [ \gamma_{1} (1-p_{1})^{2} + \gamma_{2} (1-p_{1}^{2})\right]\ee
for some constants $\gamma_1$ and $\gamma_2$  determined by the field theory.  We therefore expect:
\be c_{1} = 2 t_{b}^{-1-p_{1}} \left [ -(\gamma_{1} + \gamma_{2}) + 2 p_{1} \gamma_{1} + (\gamma_{2}-\gamma_{1}) p_{1}^{2}\right]
\label{c1expected}
\end{equation} 
The power of $t_{b}$ in the time dependence of this expected behavior matches our results, $\beta (p_{1})=-1-p_{1}$.  We may calculate $\gamma_1$ and $\gamma_2$ by setting the quadratic prefactors in \eqn{c1} and \eqn{c1expected} equal:
\begin{equation} \frac{1}{2}\left (a + b p_{1} + c p_{1}^{2}\right)= -(\gamma_{1} + \gamma_{2}) + 2 p_{1} \gamma_{1} + (\gamma_{2}-\gamma_{1}) p_{1}^{2}\end{equation}
In other words:
\begin{equation}
\begin{aligned} & \gamma_{1} = \frac{b}{4}\\
&\gamma_{2} = -\frac{1}{2} \left ( a+ \frac{b}{2}\right) = \frac{1}{2} \left ( c + \frac{b}{2}\right)  \\
\end{aligned}
\end{equation}
\noindent This is an overdetermined system and has no solution unless $a =-\left(b+c\right)$. We find that our coefficients indeed satisfy this relation and 
furthermore determine  $\gamma_{1}\approx -0.1032$ and $\gamma_{2}\approx 0.0238$. To the best of our numerical accuracy, these constants have a simple ratio $\frac{\gamma_{1}}{\gamma_{2}} =  - \frac{13}{3}$.
This is sufficient to determine the extrinsic curvature contributions to the $1/\epsilon$ divergence in general for a six dimensional bulk.

\begin{figure}
\centerline{
\includegraphics[width=0.5\textwidth]{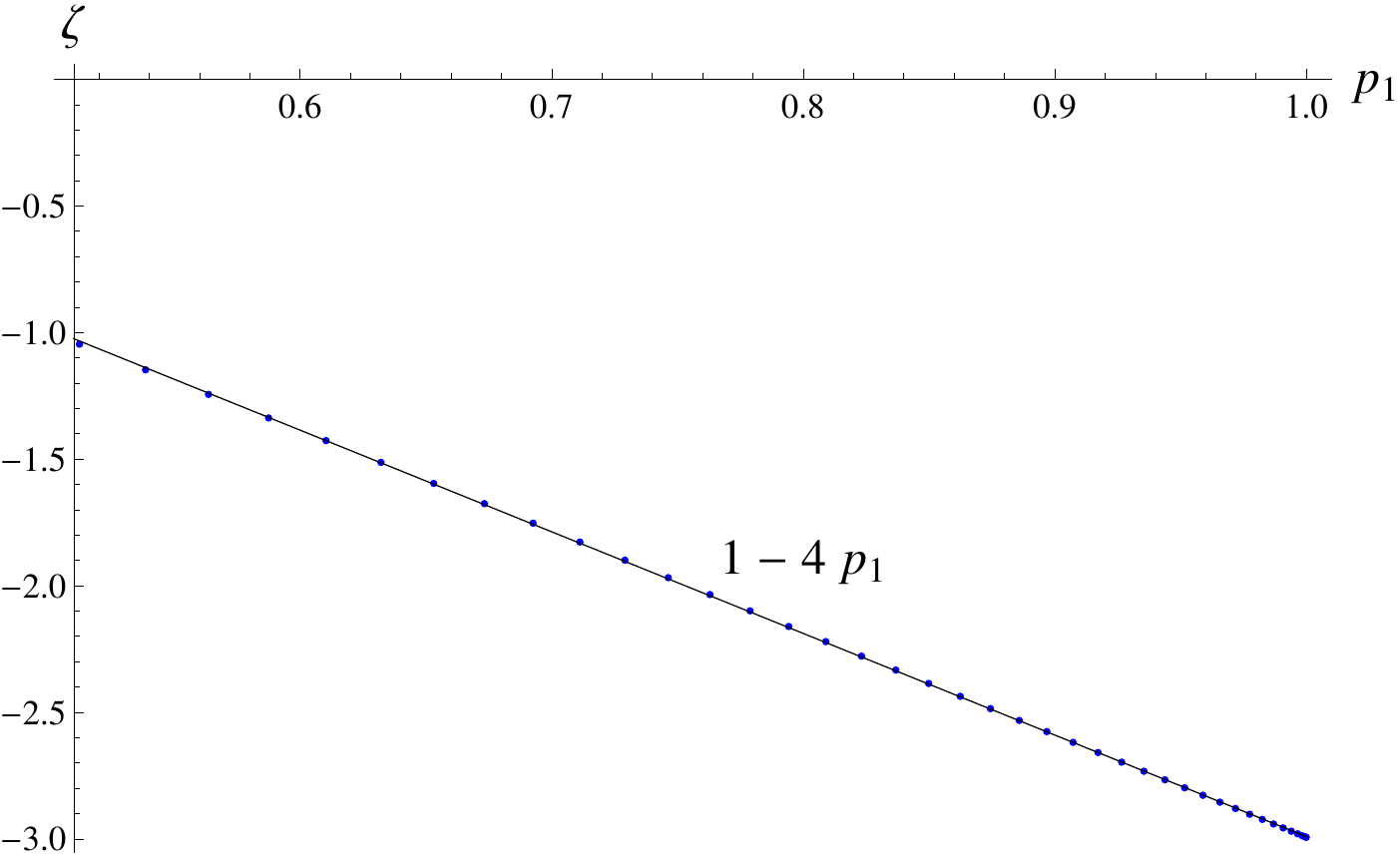}
}
\caption{\small A plot of $\zeta(p_{1})$, where $c_{0}(t) \propto t_{b}^{\zeta(p_{1})}$, for $\frac{1}{2}<p_{1}<1$. The best-fit line is $\zeta(p_{1})= 1-4p_{1}$.}
\label{fig:p1large}
\end{figure}

 We now discuss the UV finite  part of the entanglement entropy, $c_{0}(t_{b})$.  This is always negative, but has several  qualitatively different behaviors depending on the value of $p_1$.     Recall that for  a static strip in pure AdS$_{6}$,  $c_{0}$ behaves as \cite{RT09}:
\be c_{0, \ \text{static}}\propto \frac{A_{\text{bdy}}}{x_{0}^{3}} \label{C0Soliton}\ee 
\noindent A na\"{i}ve generalization of this for a time-dependent metric would be:
\begin{equation}\label{c0}
 c_{0} \propto \frac{A_{\text{bdy}}}{\mathcal{L}^{3}}
 \end{equation}
where, as before, $\mathcal{L}$ is the proper width of the entangling region. We expect to find agreement with \eqn{c0} for the Milne soliton $(p_{1}=1)$ since this is static by the coordinate transformation \eqn{MilneTransform}.  Indeed, we find that for $p_1$ between $1$ and approximately $1/2$, $c_0(t_{b})$ is well fit by a simple power law 
\begin{equation} 
c_{0} \propto t_{b}^{1-4p_{1}} = \frac{t_{b}^{1-p_{1}}}{t_{b}^{3p_{1}}} \propto \frac{A_{\text{bdy}}}{\mathcal{L}^{3}} \label{c0static} 
\end{equation} 
The agreement with this power law is illustrated in Fig.~\ref{fig:p1large}. We can attribute this agreement to the fact that for  $ 1/2 <p_1 \le 1$, the time dependence in the $x$ direction dominates over the time dependence in the $y_i$ directions.  
\noindent 

\begin{figure}
\centerline{
\includegraphics[width=0.6\textwidth]{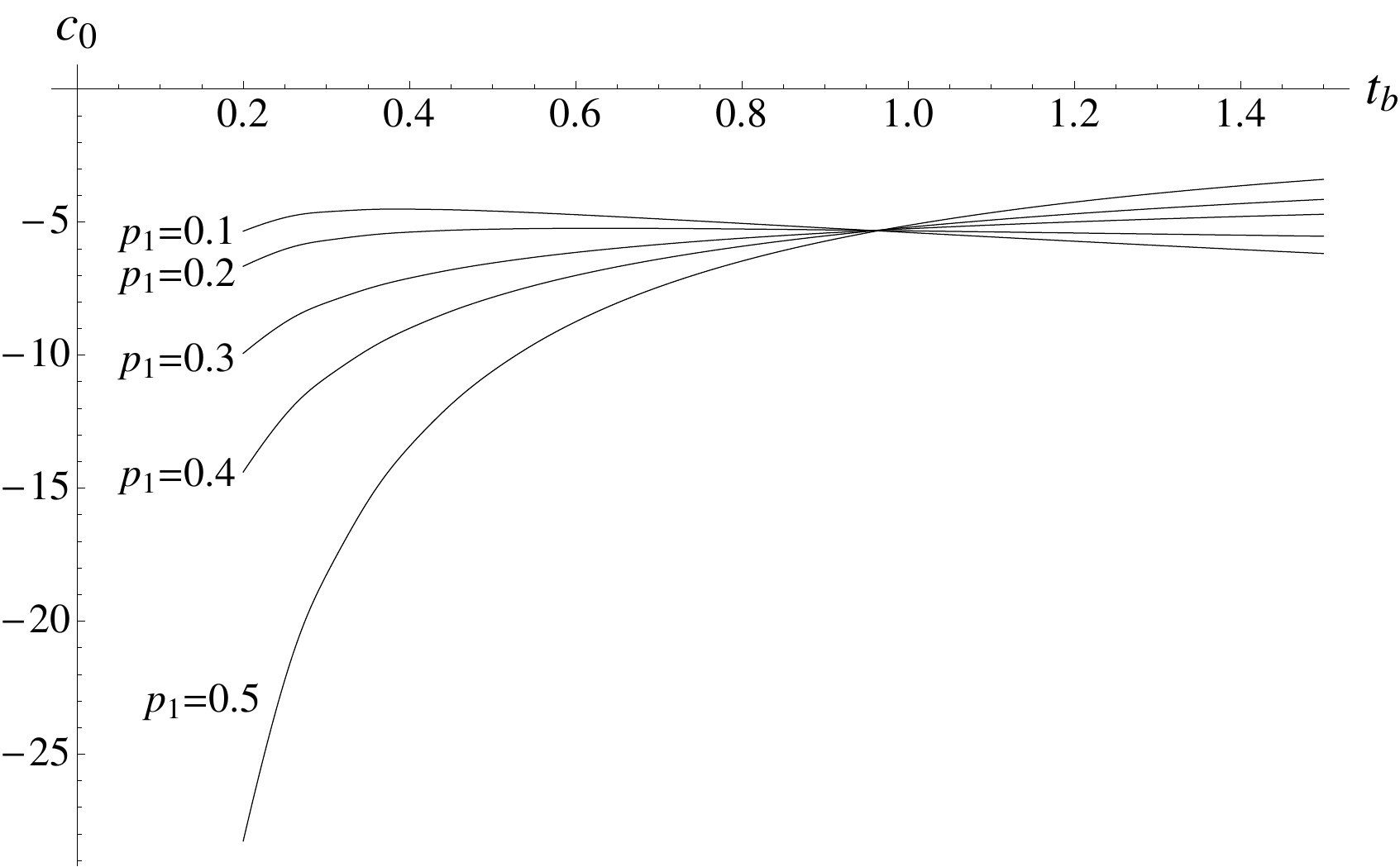}
}
\caption{\small $c_{0}$ as a function of time for $0<p_{1}<1/2$. As $p_{1}$ decreases, the $c_{0}$ curves level out, while still apparently diverging at early times, in contrast with the static prediction, which would imply that they go to zero for $p_{1}< 1/4$.}
\label{fig:p1small}
\end{figure}

If $c_{0}$ obeyed \eqn{c0static} for all $p_{1}$, there would be an abrupt change in the behavior of $c_{0}$ at the critical value $p_{1}^{(c)}=1/4$: surfaces with $p_{1}>p_{1}^{(c)}$ would have $c_{0}$ diverging at early $t_{b}$ and going to zero at late $t_{b}$, while surfaces with $p_{1}<p_{1}^{(c)}$ would have $c_{0}$ vanishing at earlier times and diverging at late times. This is not what we find. Instead, for $p_{1} < 0.5$, $c_{0}(t_{b})$ is no longer fit by a simple power law. Fig.~\ref{fig:p1small} shows the behavior of $c_{0}$ for small, positive $p_{1}$.  For $1/4 < p_1 < 1/2$, $c_{0}$ is less steep at early times than the static case would predict, while still diverging. For $0< p_1 < 1/4$, $|c_{0}(t_{b})|$ decreases as one moves toward the singularity as suggested by \eqn{c0static}, but eventually turns around and starts to increase. We believe that it will eventually diverge as $t_b \rightarrow 0$, although it is hard to numerically find the extremal surfaces very close to the singularity (and one cannot trust the classical bulk solution very close to the singularity anyway). For all of these cases, the width of the entangling region decreases at early times and the extremal surfaces stay close to the boundary. The fact that $|c_{0}(t_{b})|$ diverges at early time likely stems from
the fact that it is measuring entanglement of modes with shorter and shorter wavelength. 

It is clear from  Fig.~\ref{fig:p1small}  that all the curves cross at $t_b \approx 0.95$. In other words, $c_0(0.95)$ is essentially independent of $p_1$. This is a result of the fact that for $t\approx 1$, the Kasner metric is independent of $p_1$, and hence the same is true for the Kasner-AdS soliton. Since the extremal surfaces are staying close to a constant $t$ surface when $t\approx 1$, their area is essentially independent of $p_1$.

\begin{figure}
\centerline{
\includegraphics[width=0.5\textwidth]{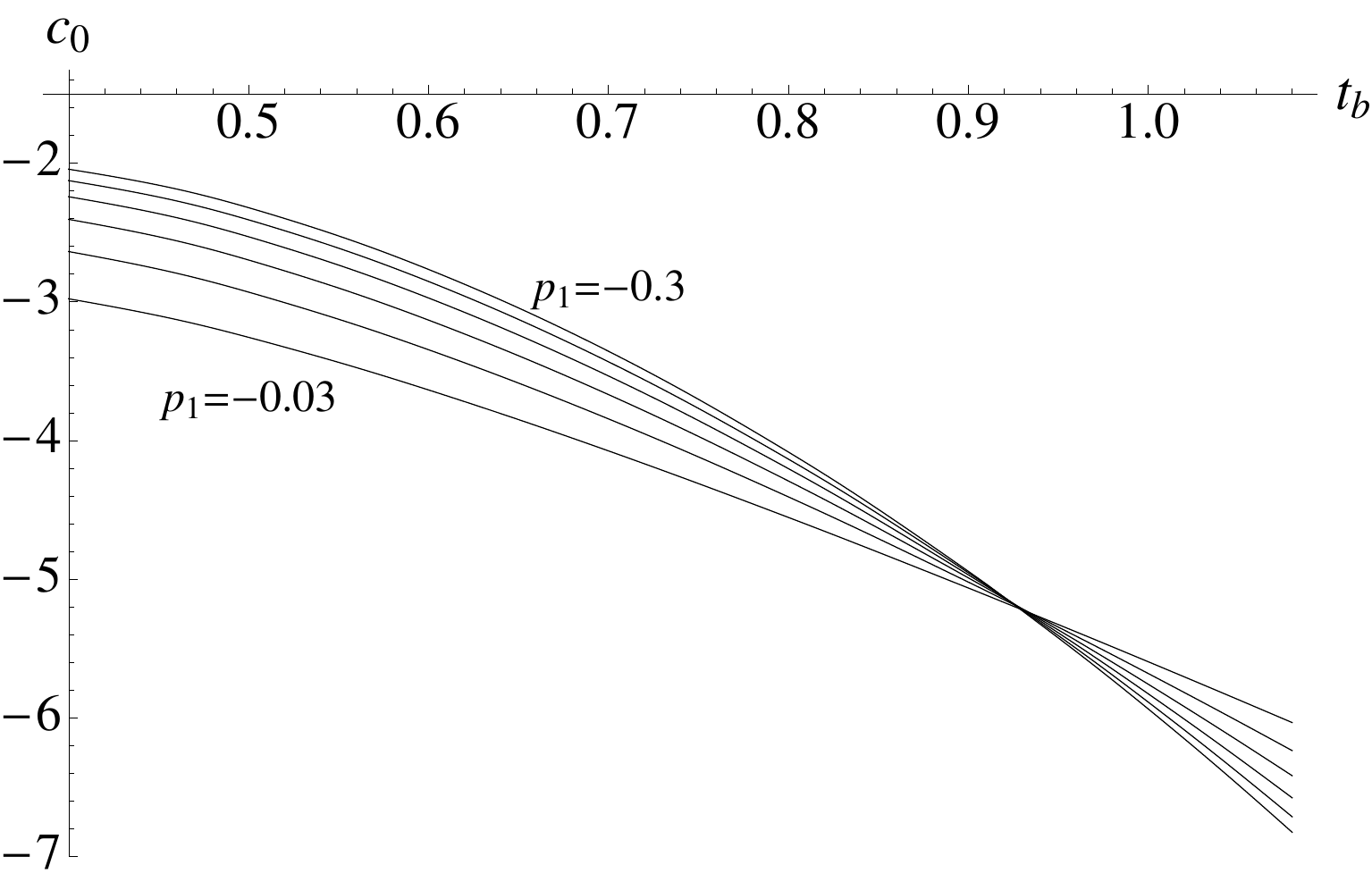}
}
\caption{\small $c_{0}$ as a function of time for $-0.3<p_{1}<0$. The curves correspond to $p_{1}=-0.03$, $-0.11$, $-0.17,$, $-0.22$, $-0.26$, and $-0.30$.}
\label{fig:p1negative}
\end{figure}

  Finally,  for $p_{1}<0$,  $|c_{0}(t_{b})|$ monotonically decreases at early times as illustrated in Fig.~\ref{fig:p1negative}. In this case, the proper width of the entangling region grows and the extremal surface probes progressively deeper into the bulk. 
   This is expected to continue until the confinement/deconfinement transition occurs, when
  the extremal surface hits the cap of the soliton and splits into two disconnected surfaces.
  
To summarize, we have computed the entanglement entropy of a confining gauge theory in a background spacetime which is a product of Kasner and a circle. In addition to the leading divergence which is proportional to the area, we found a subleading divergence proportional to the square of the extrinsic curvature of the entangling region. Near the singularity, we find that the behavior of the UV finite part of the entropy, $c_0$, can differ significantly from the na\"{i}ve extrapolation of the static result:  the behavior of the width of our entangling strip is more important than the behavior of the area of its boundary. If the width goes to zero at early time, then  $c_0$ diverges, while if it grows, then $|c_0(t_{b})|$ decreases near the singularity and eventually reaches a confinement/deconfinement transition. This shows that the approach to the singularity can be used to determine the energy scale at which a confinement/deconfinement phase transition occurs in a gapped field theory.

In cosmology, one often considers states with nonzero temperature, and a confinement/ deconfinement transition occurs at a time when the temperature reaches a critical value. In  contrast, we are not working with a state with nonzero temperature.  Our state  is given implicitly by our choice of bulk solution (since different states in the dual field theory correspond to different bulk geometries). However, one cannot describe this state as a vacuum state since the time dependence of the background causes particle creation. A vacuum state at one moment of time will not remain the vacuum at later time. This particle creation is contributing to $c_0(t_{b})$.
 
It would be interesting to compare our calculation of the entanglement entropy, which is appropriate for a strongly coupled confining theory, with the entanglement entropy of a free field in Kasner (see, e.g., \cite{Casini:2009sr}).

\vskip 2cm
\centerline{\bf Acknowledgements}
\vskip 1cm
\noindent We would like to acknowledge helpful discussions with Sebastian Fischetti, William Kelly, Rob Myers, Jorge Santos, Omid Saremi, and Benson Way. This work is supported in part by the National Science Foundation Graduate Research Fellowship under Grant No. DGE-1144085, and by NSF Grant No. PHY12-05500.

\end{document}